\documentclass[12pt]{article}
\usepackage{amssymb,amsmath,epsfig}

\begin{document}
\allowdisplaybreaks
\title{\bf Shadow of a Charged Rotating Non-Commutative Black Hole}
\author{M. Sharif \thanks{msharif.math@pu.edu.pk}~\thanks{Fellow, Pakistan Academy of Sciences,
3 Constitution Avenue, G-5/2, Islamabad.}  and Sehrish Iftikhar
\thanks{sehrish3iftikhar@gmail.com}~\thanks{On leave from Department of Mathematics, Lahore College
for Women University, Lahore-54000, Pakistan.}\\
Department of Mathematics, University of the Punjab,\\
Quaid-e-Azam Campus, Lahore-54590, Pakistan.}
\date{}
\maketitle
\begin{abstract}
This paper investigates the shadow of a charged rotating
non-commutative black hole. For this purpose, we first formulate the
null geodesics and study the effects of non-commutative charge on
the photon orbit. We then explore the effect of spin, angle of
inclination as well as non-commutative charge on the silhouette of
the shadow. It is found that shape of the shadow deviates from the
circle with the decrease in the non-commutative charge. We also
discuss observable quantities to study the deformation and
distortion in the shadow cast by the black hole which decreases for
small values of non-commutative charge. Finally, we study the
shadows in the presence of plasma. We conclude that the
non-commutativity has a great impact on the black hole shadow.
\end{abstract}
\textbf{Keywords:} Non-commutative geometry; Black hole; Shadow.\\
\textbf{PACS:} 95.30.Sf; 04.70.-s.; 04.62.+v.

\section{Introduction}

Black holes (BHs) are the end products of a complete gravitational
collapse of the massive star surrounded by a boundary from which
nothing can escape even light. Despite their dark nature, BHs exist
in the environment of the brightest objects of the universe. Black
hole does not shine but the accreting gas that spirals around, emits
radiations and then disappears when reaches the event horizon.
Generally, a BH casts a shadow in an emitting medium which arises
due to the bending of light that originates near the horizon and
travels through the accretion flow. The shape of a shadow depends on
the photon capture sphere resulting from strong bending of light
perceived by a distant observer. The shadow can actually depict the
compact object (BH or naked singularity) as its appropriate
observations can provide information about the parameters as well as
spacetime geometry around the compact object.

Black holes are usually of two types: stellar mass BHs that are
observed when they accrete matter from X-ray binaries and the
supermassive BHs in the center of galaxies. The direct observation
of BH is not possible, therefore, event horizon is the defining
characteristic which casts a relatively large shadow with respect to
a distant observer. The interest in this topic raised with the
possibility of viewing the shadow of Sgr A*, a central supermassive
BH candidate of the Milky way. It is shown that the shadow of Sgr A*
is observable with very long baseline interferometry at
submillimeter wavelengths with the assumption of thin accretion flow
in related region \cite{1}. Further developments in this framework
can be viewed in the literature \cite{2}.

The concept of a BH shadow initiated with Bardeen \cite{3} as well
as Cunningham and Bardeen \cite{4} who studied shadow of the Kerr
BH. This technique can also be found in \cite{5}, where coordinates
of the observer's sky have been calculated. Different aspects
related to shadows for the Kerr-Newman BH were discussed in
\cite{6}. Hioki and Maeda \cite{7} discussed that BH can be
distinguished from naked singularity by examining the shadows of
Kerr BH as well as naked singularity. They provided a technique to
measure the spin and angle of inclination by defining two
observables. The properties of different BHs shadows have been
extensively studied in the literature \cite{8}. The shadow of binary
BHs is another interesting subject. Bohn et al. \cite{9} showed that
the shadow of binary BHs is different from the superposition of two
singleton BH shadows. Shipley and Dolan \cite{10} explored the
shadows of two extremely charged binary BHs and stated that they are
interesting examples of chaotic scattering in nature. Another
fascinating feature of the BH shadows is that it can serve to test
the no hair theorem \cite{11}.

The occurrence of curvature singularities in gravitational collapse
has led to important developments in general relativity (GR). To
resolve the problem of singularities, quantum gravity proved to be
very successful. Since GR breaks down at short distances, the
non-commutative (NC) theory provides a useful approach to discuss
the spacetime dynamics at Planck scale. Gross and Mende \cite{12a}
explored the behavior of string interactions at short distances.
Maggiore \cite{12b} found generalized uncertainty principle which is
consistent with the results obtained in string theory. He suggested
that the minimum length (of order Planck length) emerges naturally
from any theory of quantum gravity. The structure of NC geometry is
based on universal minimal length scale which is equivalent to
Planck's length. In this framework, the spacetime quantization
requires that its coordinates become Hermitian operators which do
not commute, i.e., \cite{12}
\begin{equation*}
[x^{A},x^{B}]=\iota \Theta^{AB},
\end{equation*}
where $\Theta^{AB}$ is a real-valued antisymmetric matrix with
$\Theta^{AB}=\Theta \text{diag}(\epsilon_{ij},~\epsilon_{ij},...)$,
$\Theta$ is a constant of dimension $\text{(lenght)}^{2}$. In the
limit $\Theta\rightarrow 0$, the ordinary spacetime is recovered.
Physically, $\Theta^{AB}$ represents a small patch in $AB$-plane of
the observable area as the Planck's constant ($\hbar$) illustrates
the smallest fundamental cell of the observable phase space in
quantum mechanics.

The quantum gravity BHs incorporates the effects of quantum gravity
in the vicinity of origin, where the classical curvature singularity
appears. Many singularity free BHs have been obtained in the
framework of quantum gravity such as BHs inspired by NC geometry. In
the early 90's, Doplicher et al. \cite{14a,14aa} provided a detailed
study about the spacetime non-commutativity which shows that GR
together with the uncertainty principle of quantum mechanics leads
to a class of models of NC spacetime. There are some comprehensive
reviews about NC field theory and the spacetime with NC coordinates
\cite{14b}. The NC form of different BHs such as charged \cite{14},
higher dimensional \cite{15}, rotating \cite{16} have also been
obtained. The effects of non-commutativity on the properties of BHs
have extensively been studied. Nicolini et al. \cite{17}
investigated radiating NC Schwarzschild BH and found that
non-commutativity can cure singularity problems. Nozari and
Mehdipour \cite{18} explored quantum tunneling from NC Schwarzschild
BH and showed that the information would be preserved by a stable BH
remnant due to NC effects. The lensing properties of NC BHs have
also been explored. Ding et al. \cite{19} studied the influence of
NC parameter on the strong field lensing of NC Schwarzschild BH and
found that the NC parameter has a similar behavior as that of charge
of Reissner-Nordstr$\ddot{o}$m BH. Wen-Wei \cite{20} found that the
NC parameter affects the shape of BH shadow and increases its
deformation as compared to Kerr BH. Many other important aspects of
spacetime non-commutativity have been discussed in literature
\cite{21}.

Generally, BHs are characterized by non-zero angular momentum which
breaks the spherical symmetry resulting in rotational symmetry
around axis. The gravitational field around BH is highly affected by
its rotation, therefore, it is natural to expect some important
effects on the shadow cast by the BH. It is observed that the shadow
of non-rotating BH has circular shape while the shadow of rotating
BH is deformed due to its spin \cite{5,7,23}. Another motivation
comes from the fact that Sgr A* - the most remarkable candidate for
BH which likely have spin whose axis is perpendicular to the
galactic plane. The predicted size of the shadow of Sgr A* is $\sim$
30 $\mu$ arcseconds \cite{1}. According to no hair theorem, any
stationary, asymptotically flat BH vacuum solution of the field
equations can be characterized by three variables: mass, charge and
angular momentum per unit mass ($M,~Q,~a$). This leads to the
formation of Kerr-Newman (KN) BH as an outcome of non-spherical
gravitational collapse. It is suggested that an isolated rotating BH
cannot have an electromagnetic field except it endowed with a net
electric charge \cite{c1}. It is possible that Kerr Newman BH could
be the end point of magnetized, massive rotating stars as a result
of catastrophic gravitational collapse \cite{c2}. It is also
proposed that the charged rapidly rotating galactic BHs (Kerr Newman
BHs) emit gamma rays (in a bipolar out flow) \cite{c3}.

The effect of plasma on the properties of BH has extensively been
studied. Muhleman and Johnston \cite{26} discussed the deflection of
light rays by gravity as well plasma by considering radio waves
traveling near the Sun. He used a product of the plasma and
gravitational refractive indices in weak field approximation. Breuer
and Ehlers \cite{27} explored the Hamiltonian for light rays in a
magnetized pressureless plasma. A similar derivation was found by
Perlick \cite{28} for the pressureless, non-magnetized plasma.
Perlick et al. \cite{30} discussed the effect of plasma on the
shadow of Ellis wormholes as well as spherically symmetric BH and
found that the decrease in the shadow by the presence of plasma.
Rogers \cite{c} explored gravitational lensing and circular orbits
of the Schwarzschild BH for plasma density distributions $N=1/r^{h}$
with $h\geq0$. There is a large body of literature available
\cite{31} on the effect of plasma on the shadows cast by the BHs.

In this paper, we study the shadow of charged rotating NC BH in the
equatorial as well as non-equatorial plane. The paper is organized
as follows. In the next section, we discuss the null geodesics for
the NC BH. Section \textbf{3} is devoted to study the apparent shape
of the NC geometry inspired shadows and the behavior of observables.
In section \textbf{4}, we explore the influence of plasma on the
shape of shadows. The results are concluded in the last section.

\section{Geodesics in a Non-Commutative Spacetime}

Non-commutativity is an intrinsic property of the spacetime which
implies the existence of a natural ultra-violet cutoff or
equivalently a minimal (measurable) length in quantum field theory.
The NC spaces have a long history. The first model of NC spacetime
(admitting a minimal length scale and invariant under the Lorentz
transformation) was proposed by Snyder \cite{aa} which was an
attempt to introduce a short distance cutoff in quantum field
theory. The interest in NC theory raised with the seminal paper of
Seiberg and Witten \cite{aa1} in which they extended NC theory to
the string theory with a constant $B$-field. Some important examples
of non-commutativity includes Lei algebra type deformation (the most
studied example is $\kappa$-Minkowski spacetime \cite{aa2}) and
canonical type deformation (a simple and widely discussed example of
NC spacetime) \cite{14a,aa1,aa3}. In NC geometry, the pointlike
smeared gravitating source
($\rho_{\Theta}(r)=\frac{1}{(4\Theta)^{d/2}}e^{-x^2/4\Theta},~d$ is
the dimension of spacetime) is assumed to remove divergencies in GR.
Modesto and Nicolini \cite{22} obtained regular charged rotating BH
solution in the background of NC geometry (based on canonical
deformation) by employing Newman-Janis mechanism. They found that
the resulting BH solution is unique and exhibits local regular
behavior at the origin as well as standard rotating geometries at
large distances. The Ricci scalar as well as Kretschmann invariant
remain finite at the origin. The charged NC BH in Boyer-Lindquist
coordinates is given as
\begin{eqnarray}\nonumber
ds^{2}&=&(\frac{\Delta-a^2\sin^2\theta}{\Sigma})dt^2
-\frac{\Sigma}{\Delta}dr^2-\Sigma d\theta^2 \\\nonumber
&+&2a\sin^2\theta [1-\frac{\Delta-a^2\sin^2\theta}{\Sigma}]dt d\phi
\\\label{1}
&-&\sin^2\theta[\Sigma+a^2\sin^2\theta(2-\frac{\Delta
-a^2\sin^2\theta}{\Sigma})]d\phi^2,
\end{eqnarray}
where
\begin{eqnarray*}
&&\Delta=r^2+a^2-2m(r)r+q^2(r),\quad\Sigma=r^2+a^2\cos\theta^2,\\
&&m(r)=\frac{M\gamma(\frac{3}{2};\frac{r^2}{4\Theta})}{\Gamma(\frac{3}{2})},\\
&&q^2(r)=\frac{Q^2}{\pi}[\gamma^2(\frac{1}{2};\frac{r^2}{4\Theta})
-\frac{r}{\sqrt{2\Theta}}\gamma(\frac{1}{2};\frac{r^2}{2\Theta})
+r\sqrt{\frac{2}{\Theta}}\gamma(\frac{3}{2};\frac{r^2}{4\Theta})].
\end{eqnarray*}
The event horizon of (\ref{1}) is the largest root of $\Delta=0$
given by
\begin{equation}\nonumber
r^2+a^2-2m(r)r+q^2(r)=0.
\end{equation}
A detailed discussion about the nature of horizons and
thermodynamical properties are available in \cite{22}. Here we would
like to discuss the horizons briefly. The function $\Delta=0$
produces horizons for $M\geq M_{extr}$, i.e., for extremal and
non-extremal cases, (\ref{1}) represents a BH but for $M<M_{extr}$,
there is no horizon and the metric (\ref{1}) describes a regular
charged spinning object such as charged spinning mini gravastar
\cite{22}. In this paper, we do not consider horizonless case. For
this purpose, we find maximum values for the spin and charge
parameter corresponding to different values of non-commutative
parameter by solving $\Delta(r)=0$ and $\frac{d\Delta(r)}{dr}=0$
numerically (Figure \textbf{1}).

The particle motion can be described by the Lagrangian
\begin{figure}\centering
\epsfig{file=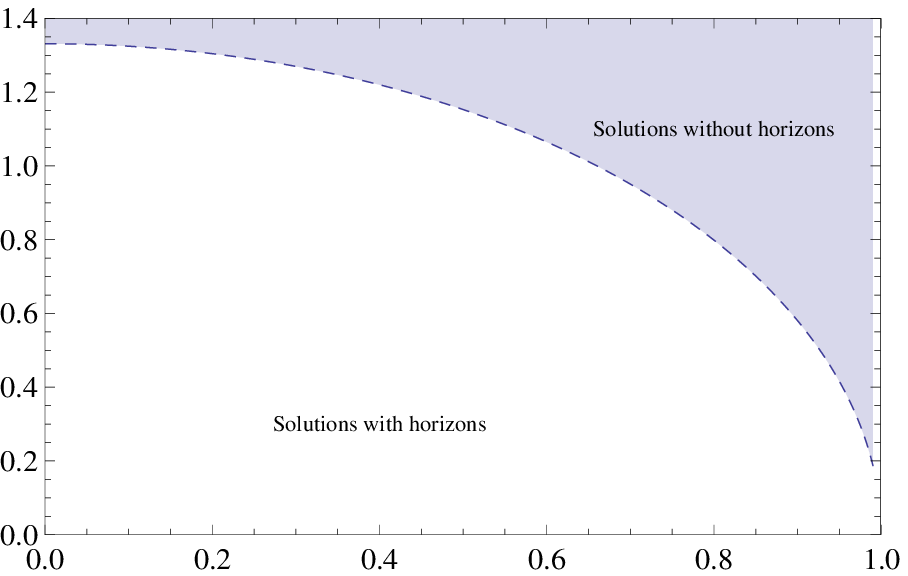,width=.43\linewidth}\epsfig{file=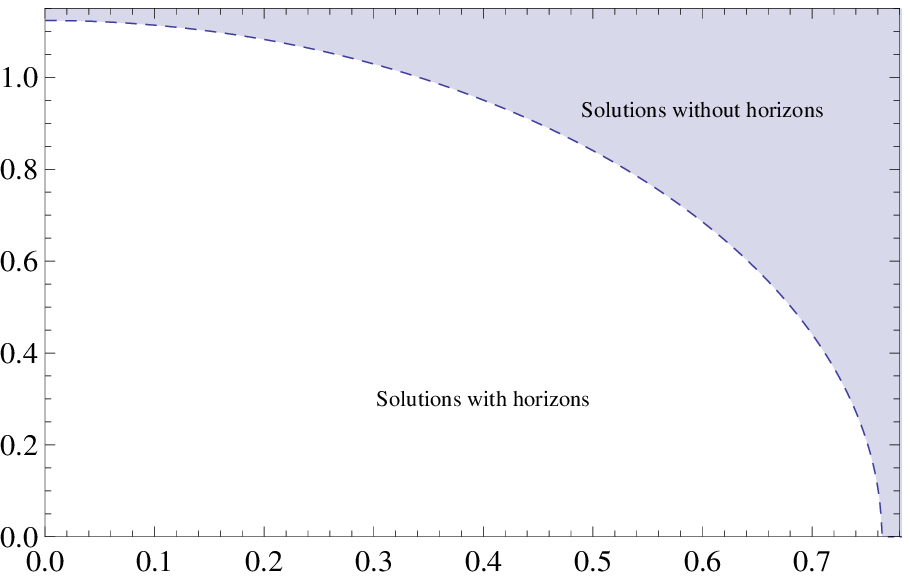,width=.43\linewidth}
\\\epsfig{file=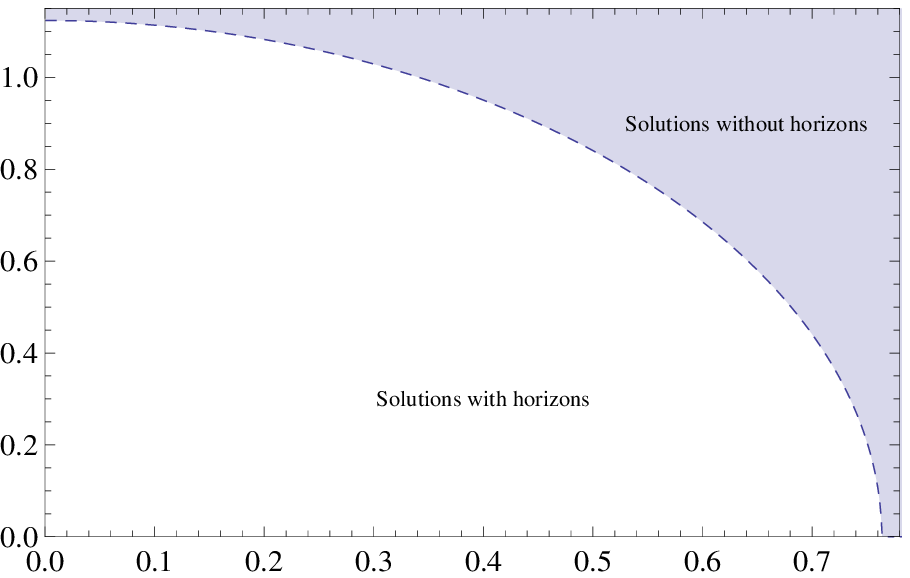,width=.43\linewidth}\epsfig{file=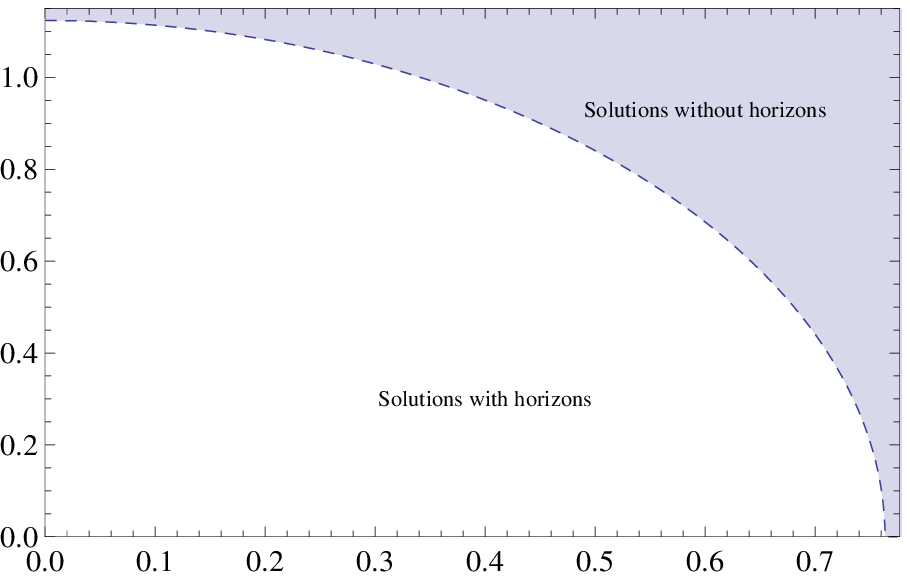,width=.43\linewidth}
\caption{The parameter space indicates the charged rotating NC BH.
Here, the vertical axis shows $Q/M$ and the horizontal axis
represents $a/M$. The blue line is the boundary that separates the
region of BH from non-BH. The upper panel plots correspond to
$\Theta=0.1$ (left), $\Theta=0.2$ (right) while lower panel to
$\Theta=0.3$ (left), $\Theta=0.4$ (right).}
\end{figure}
\begin{equation}\label{r}
\mathcal{L}=\frac{1}{2}g_{\nu\sigma}\dot{x}^{\nu}\dot{x}^{\sigma},
\end{equation}
where $\dot{x}^{\nu}=u^{\nu}=dx^{\nu}/d\lambda$, $u^{\nu}$ is the
particle's $4$-velocity and $\lambda$ is the affine parameter. The
energy $E$ and angular momentum $L$ are given by
\begin{equation}\label{r}
E=p_{t}=\frac{\partial \mathcal{L}}{\partial \dot{t}}=g_{\phi
t}\dot{\phi}+g_{tt}\dot{t}, \quad L=-p_{\phi}=-\frac{\partial
\mathcal{L}}{\partial \dot{\phi}}=-g_{\phi \phi}\dot{\phi}-g_{\phi
t}\dot{t},
\end{equation}
The Hamilton-Jacobi equation yields
\begin{eqnarray}\label{9}
\frac{\partial S}{\partial
\lambda}=\frac{1}{2}g^{\nu\sigma}\frac{\partial S}{\partial
x^{\nu}}\frac{\partial S}{\partial x^{\sigma}}.
\end{eqnarray}
We find that the Lagrangian is independent of $t$ and $\phi$,
therefore $p_{t}$ and $p_{\phi}$ are conserved and hence describe
stationary and axisymmetric characteristics of charged rotating NC
BH. Equation (\ref{9}) takes the form
\begin{eqnarray}\label{12}
S=\frac{1}{2}m^2_{0}\lambda-Et+L\phi+S_{r}(r)+S_{\theta}(\theta).
\end{eqnarray}
For the metric (\ref{1}), it leads to
\begin{eqnarray}\label{3n}
\Sigma\frac{\partial
t}{\partial\lambda}&=&a(L-aE\sin^2\theta)+\frac{r^2+a^2}{\Delta}[E(r^2+a^2)-aL],
\\\label{6n}
\Sigma\frac{\partial
\phi}{\partial\lambda}&=&(L\csc^2\theta-aE)+\frac{a}{\Delta}[E(r^2+a^2)-aL],
\\\label{4n1}
\Sigma\frac{\partial r}{\partial\lambda}&=&\pm\sqrt{R},
\\\label{4n}
\Sigma\frac{\partial \theta}{\partial\lambda}&=&\pm\sqrt{\Theta},
\end{eqnarray}
where
\begin{eqnarray}\label{14}
R&=&[E(r^2+a^2)-aL]^2-\Delta[K+(L-aE)^2],\\\label{15}
\Theta&=&K+\cos^2\theta(a^2E^2-L^2\csc^2\theta),
\end{eqnarray}
and $K$ is the separation constant introduced by Carter \cite{31a}.
Equation (\ref{4n1}) can be represented as
\begin{equation}
(\Sigma\frac{\partial r}{\partial\lambda})^2+V_{eff}=0,
\end{equation}
where
$$V_{eff}=[E(r^2+a^2)-aL]^2-\Delta[K+(L-aE)^2].$$
The above equation can describe radial motion of the particles.
Figure \textbf{2} shows that orbits of the photons are initially
unstable for small values of $r$, it approaches to maximum values
(which shows the stable orbits) and finally becomes unstable with
increasing $r$. We also note that stable points for the photons in
the charged NC geometry corresponding to large values of charge are
greater than that of small values of charge.
\begin{figure}\centering
\epsfig{file=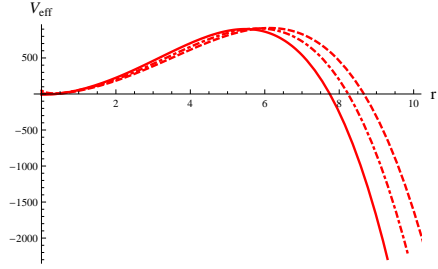,width=.42\linewidth}\caption{Plots of the
effective potential versus $r$ corresponding to $Q=0.2$ (solid),
$Q=0.3$ (dashed) and $Q=0.5$ (dot-dashed) with $\Theta=0.3$ and
$a=0.2$.}
\end{figure}

The conditions for unstable circular orbit $R(r)=0$ and
$\frac{dR(r)}{dr}=0$ \cite{5} lead to
\begin{eqnarray}\label{14.1}
&&r^4+(a^2-\xi^2-\eta)r^2+(2m(r)r-q^2(r))(\eta+(\xi-a)^2)-a^2\eta=0,
\\\nonumber
&&4r^3+2(a^2-\xi^2-\eta)r+(2m'(r)r+2m(r)-2q(r)q'(r))\\\label{14.2}
&&\times(\eta+(\xi-a)^2)=0,
\end{eqnarray}
where $\xi=L/E$, $\eta=K/E^2$ are the impact parameters and
\begin{eqnarray}\nonumber
m'(r)&=&-\frac{e^{-\frac{r^2}{4 \Theta}} M r
\sqrt{\frac{r^2}{\Theta}}}{2 \Theta \sqrt{\pi}},\\\nonumber
q'(r)&=&\frac{Q^2}{\pi}[\frac{e^{\frac{r^2}{2\Theta
r^2}}}{\Theta^{3/2} \sqrt{\frac{r^2}{\Theta}}} -
\frac{\sqrt{2}r^2e^{\frac{r^2}{2\Theta r^2}}}
{\Theta^{3/2}\sqrt{\frac{r^2}{\Theta}}} + \sqrt{2}
\sqrt{\frac{1}{\Theta}} \gamma(\frac{1}{2}, \frac{r^2}{4 \Theta})
\\\nonumber
&-& \frac{2r e^{\frac{r^2}{2\Theta r^2}}\gamma(\frac{1}{2},
\frac{r^2}{4 \Theta})}{\Theta \sqrt{\frac{r^2}{\Theta}}} -
\frac{\gamma(\frac{1}{2}, \frac{r^2}{2 \Theta})}{\sqrt{2\Theta}}].
\end{eqnarray}
Solving Eqs.(\ref{14.1}) and (\ref{14.2}), we have
\begin{eqnarray}\nonumber
\xi&=&\frac{1}{a[m(r)+r(m'(r)-1)-q(r)q'(r)]}
[m(r)(a^2-3r^2)+r(r^2+a^2)
\\\label{17}
&\times&(m'(r)+1)+2q^2(r)-q(r)q'(r)(r^2+a^2)],
\\\nonumber
\eta&=&\frac{r^2}{a^2[m(r)+r(m'(r)-1)]^2}[m(r)(4a^2-9m(r)r+6r^2)r-2m'(r)
\\\nonumber
&\times&(2a^2+r^2-3m(r)r)r^2-r^4(m'^2(r)+1)-4q^2(r)
\\\nonumber
&\times&(a^2+q^2(r)-3m(r)r+m'(r)r^2+r^2)-q'(r)(4a^2+4q^3(r)
\\\label{17aa}
&-&6m(r)q'(r)r-2m'(r)q(r)r^2-q(r)r+2q(r)r^2)].
\end{eqnarray}
The above equations are important as they are related to the
celestial coordinates of the image seen by an observer at infinity
and determine the contour of the BH shadow (will be discussed in the
next section).
\begin{figure}\centering
\epsfig{file=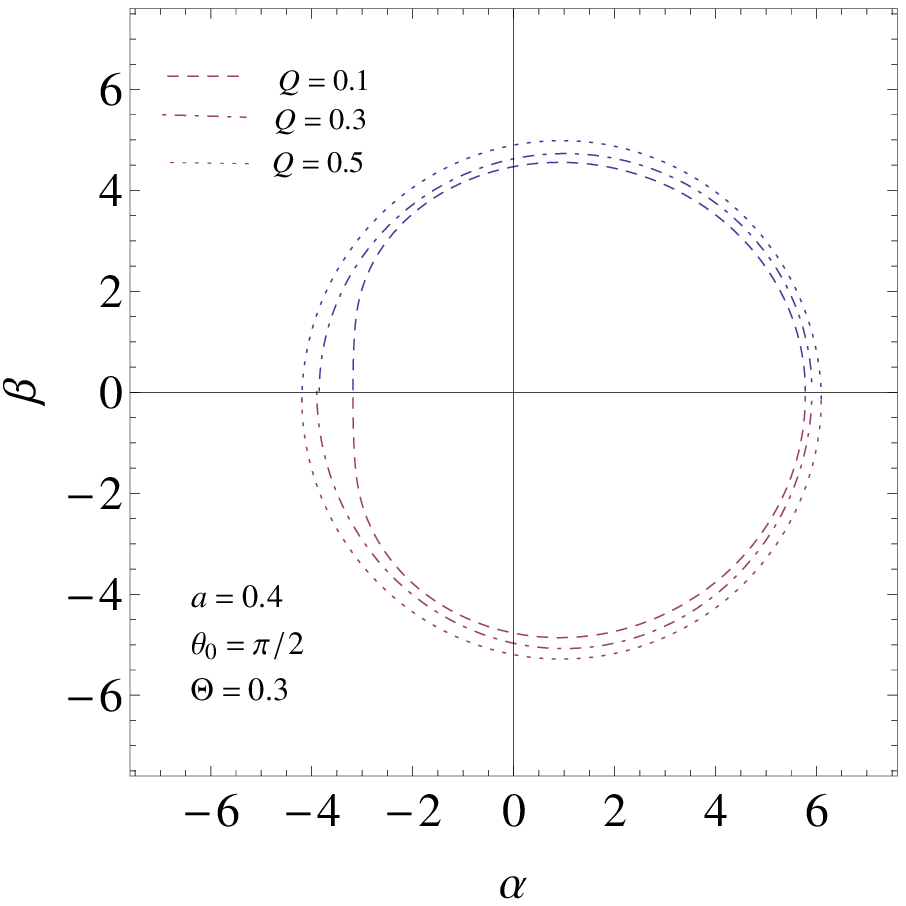,width=.43\linewidth}\epsfig{file=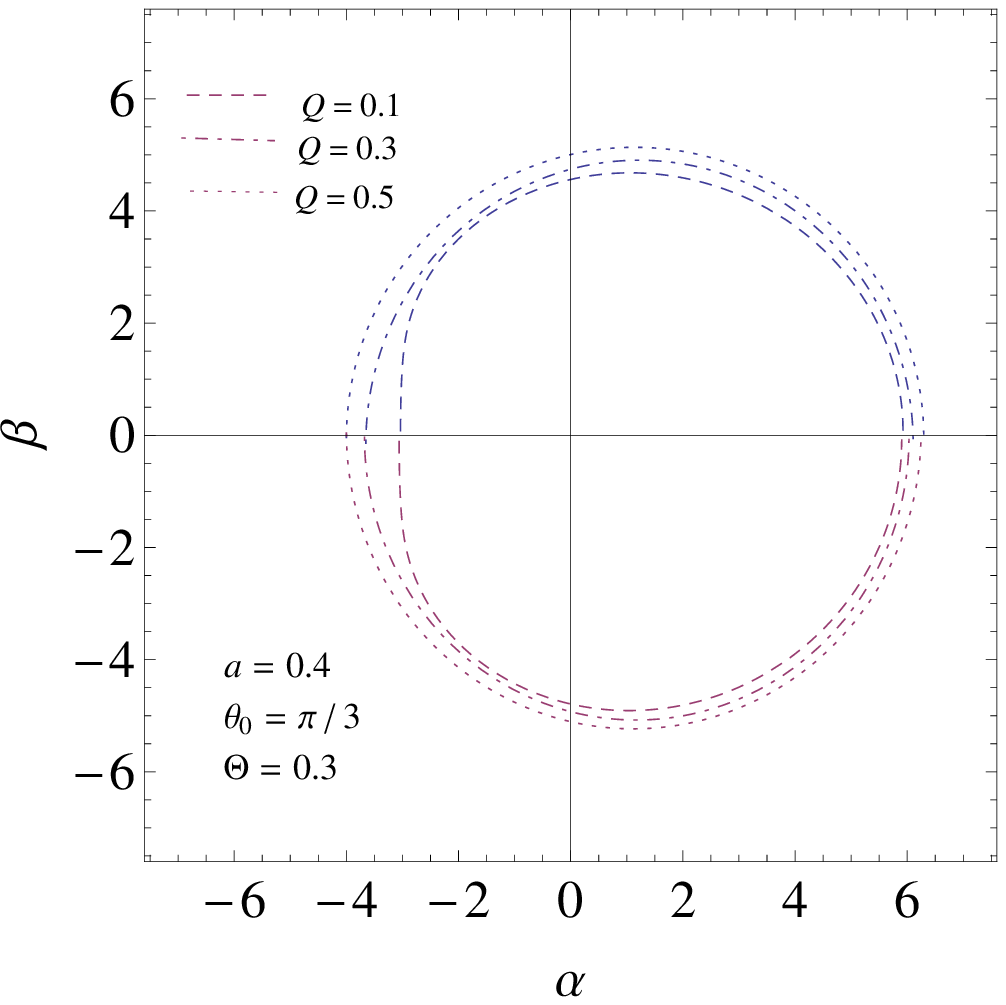,width=.43\linewidth}
\\
\epsfig{file=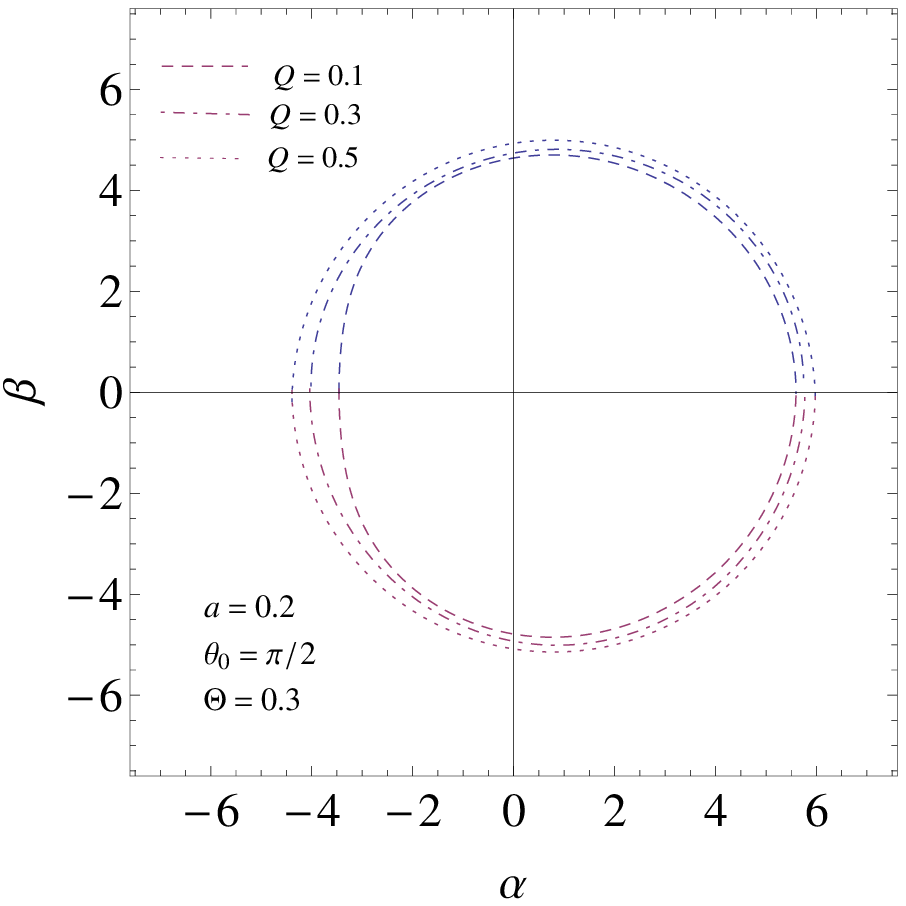,width=.43\linewidth}\epsfig{file=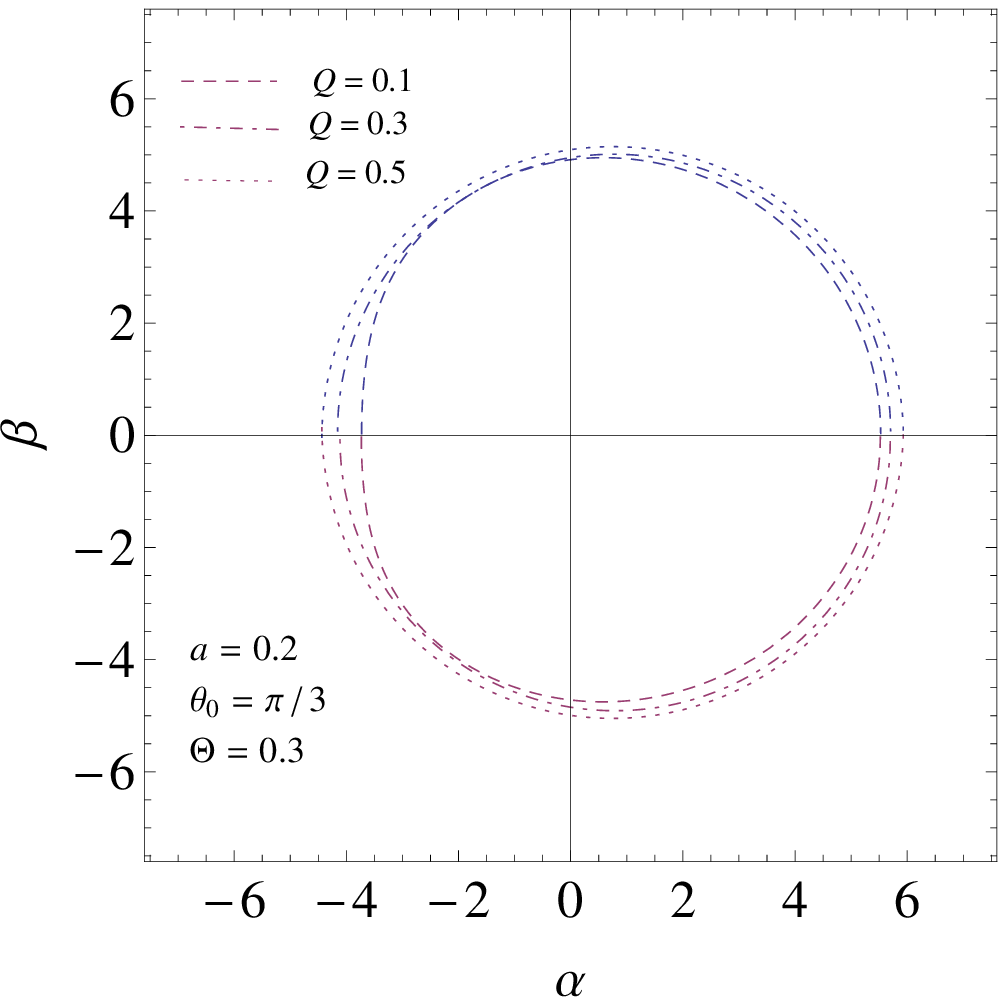,width=.43\linewidth}
\\
\epsfig{file=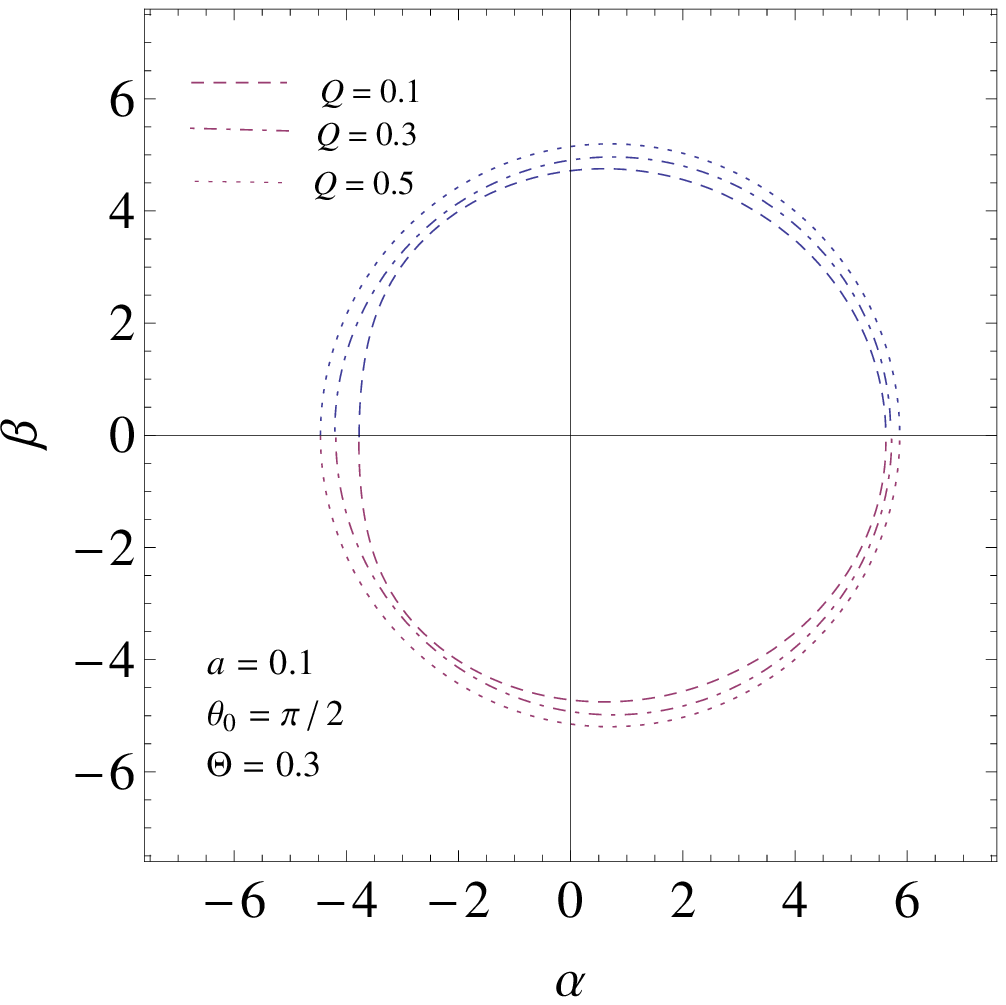,width=.43\linewidth}\epsfig{file=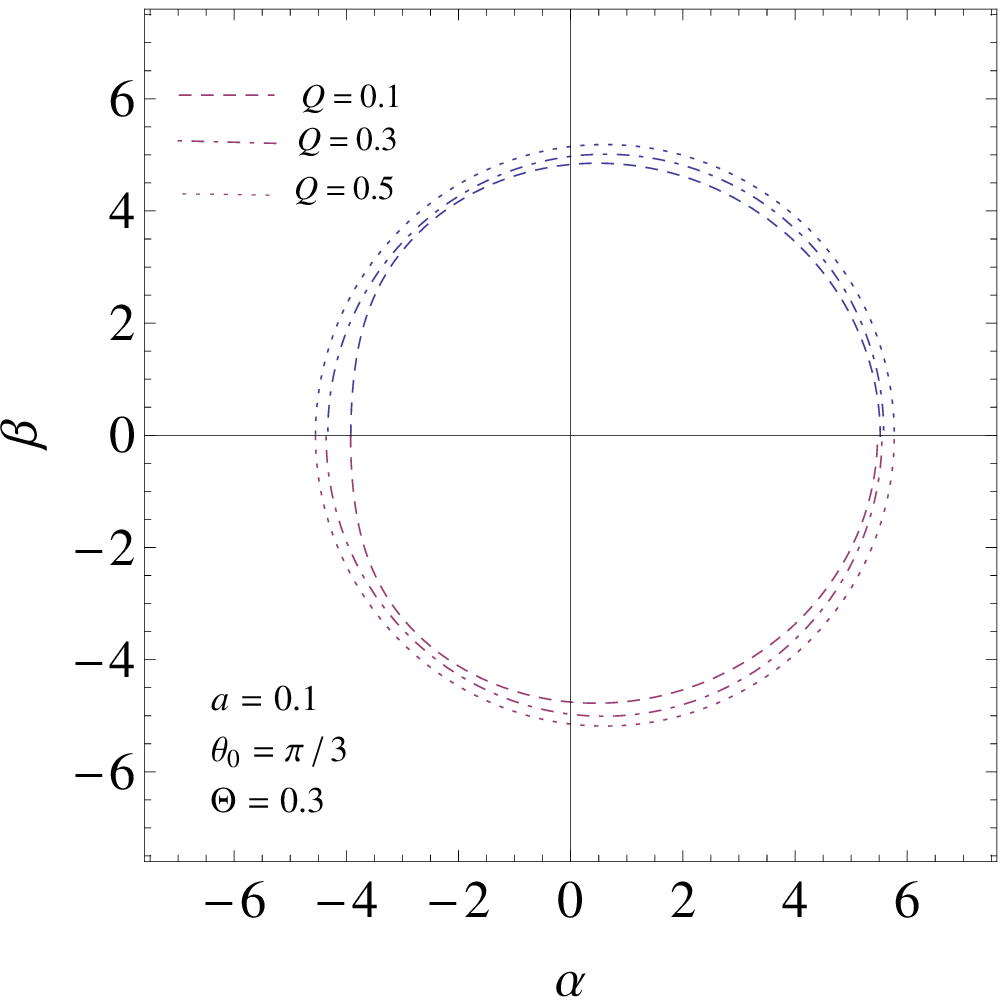,width=.43\linewidth}
\caption{Shadows of charged NC BH.}
\end{figure}

\section{Some Features of Black Holes}

Here, we study some features of BHs related to its shadows.

\subsection{Shadows}

Photons that emit from a background source of a BH suffer deflection
due strong gravitational field. The photons having small impact
parameter fall into the BH and form a dark region in sky. An
observer viewing that part of the sky would observe a black spot
having larger radius than the event horizon around the BH. This dark
region is called the shadow of the BH. The shape of the BH shadow
for a far away observer can be determined by the following celestial
coordinates \cite{23}
\begin{eqnarray}\label{18}
\alpha &=&\lim\limits_{r\rightarrow\infty}
(-r^2\sin\theta\frac{d\phi}{dr}|_{\theta\rightarrow
\theta_{0}}),\\\label{19}
\beta&=&\lim\limits_{r\rightarrow\infty}(r^2\frac{d\theta}{dr}|_{\theta\rightarrow
\theta_{0}}),
\end{eqnarray}
where $\theta_{0}$ is the angle of inclination between the rotation
axis of BH and observer's line of sight. The celestial coordinates
$\alpha$ and $\beta$ represent the apparent perpendicular distances
of the image around the BH, seen from the axis of symmetry and from
its projection on equatorial plane, respectively. These coordinates
provide the apparent position of the image in the plane which passes
through the center of BH and is orthogonal to the line joining the
BH and observer. Using Eqs.(\ref{6n})-(\ref{4n}) in
(\ref{18})-(\ref{19}), we have
\begin{eqnarray}\label{18a}
\alpha&=&-\xi\csc\theta_{0},
\\\label{19a}
\beta&=&\sqrt{\eta+a^2\cos^2\theta_{0}-\xi^2\cot^2\theta_{0}}.
\end{eqnarray}
Here $\xi$ and $\eta$ are given by Eqs.(\ref{17})-(\ref{17aa}). The
shadow of a charged rotating NC BH is shown in Figure \textbf{3}.
The silhouette of the shadow is influenced by NC mass, charge, spin
and the angle of inclination. The shape of the shadow changes with
respect to the NC charge. The high value of NC charge causes a
decrease in the shadow. For the low values of charge, the shape
deviates from the standard circle. Moreover, for the highly spinning
BH, the shadow is more deformed than the slowly spinning BH. There
is more deformation for $\theta_{0}=\pi/2$ as compared to
$\theta_{0}=\pi/3$.

\subsection{Observables}

Here, we study observables for the charged rotating NC BH. The
observables can be used to determine important astronomical
information about the collapsed objects (BHs and naked
singularities) such as to characterize the apparent shape of the
shadow. The silhouette of BH shadow determines apparent image of the
photon capture sphere seen by the observer at large distance. Hioki
and Meada \cite{7} proposed that the shadows's silhouette passes
through four points, the top $(\alpha_{t},\beta_{t})$, bottom
$(\alpha_{b},\beta_{b})$, right $(\alpha_{r},0)$ and left point
$(\alpha_{l},0)$. They presented two parameters as observable
quantities. Firstly, the radius $R_{s}$ determines the size
(approximate) of the shadow defined as the radius passing through
three different points, i.e., top, bottom and the most right points
of a reference circle. The second is $\delta_{s}$ which measures the
deformation and is defined as $\delta_{s}=\frac{D_{s}}{R_{s}}$,
where $D_{s}$ is the difference between the most left points, i.e.,
$(\alpha_{l},0)$ and $(\tilde{\alpha}_l,0)$. The point
$(\tilde{\alpha}_l,0)$ cuts the horizontal axis opposite to
$(\alpha_{r},0)$. This discussion can be explained through the
figure 1 given in \cite{24}. The plots of radius of shadow and
deformation parameter are shown in Figure \textbf{4}. We observe
that the deformation parameter has increasing behavior along spin
for the varying values of NC charge. For large value of NC charge,
the deformation gradually decreases which means that the shape of
shadow will approximately be a circle. The deformation is high for
the high values of spin. The radius of the shadow has decreasing
behavior with respect to spin.
\begin{figure}\centering
\epsfig{file=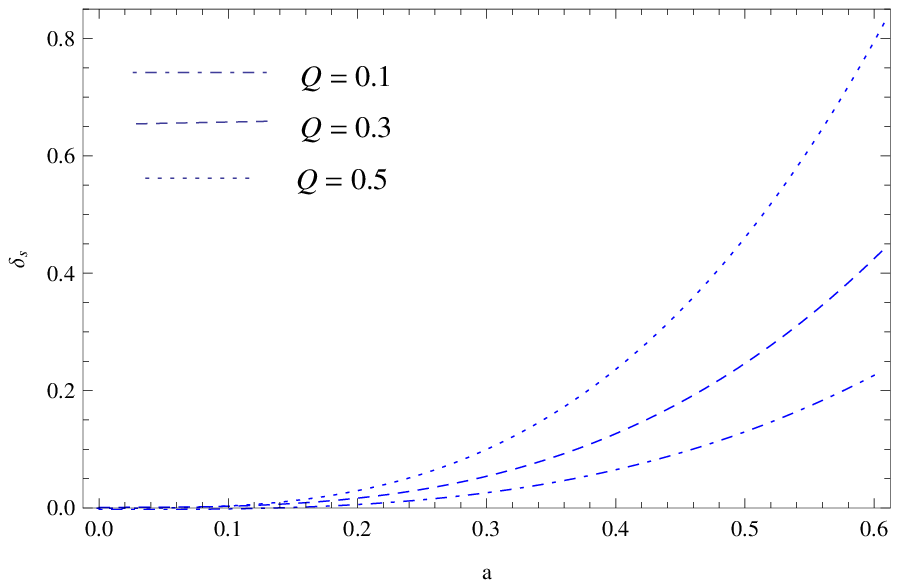,width=.40\linewidth}\epsfig{file=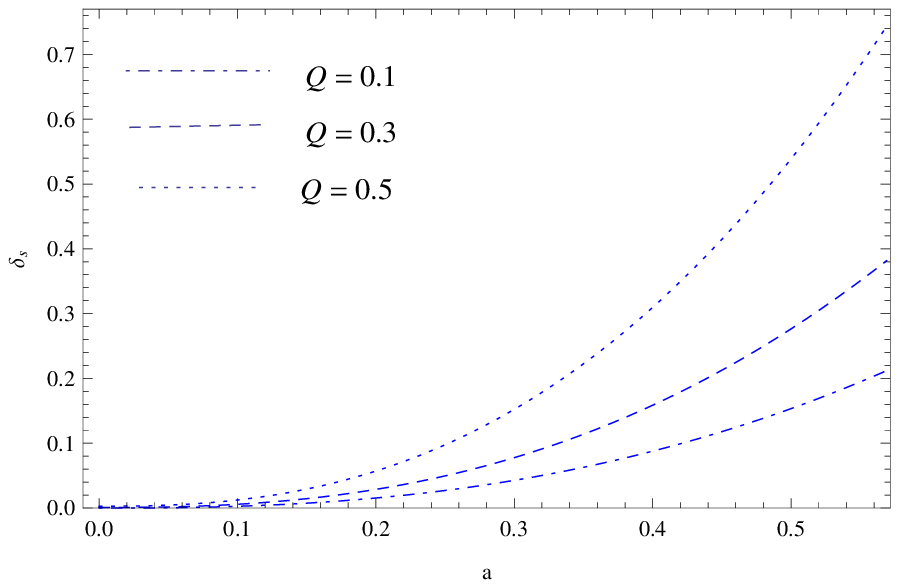,width=.40\linewidth}
\\
\epsfig{file=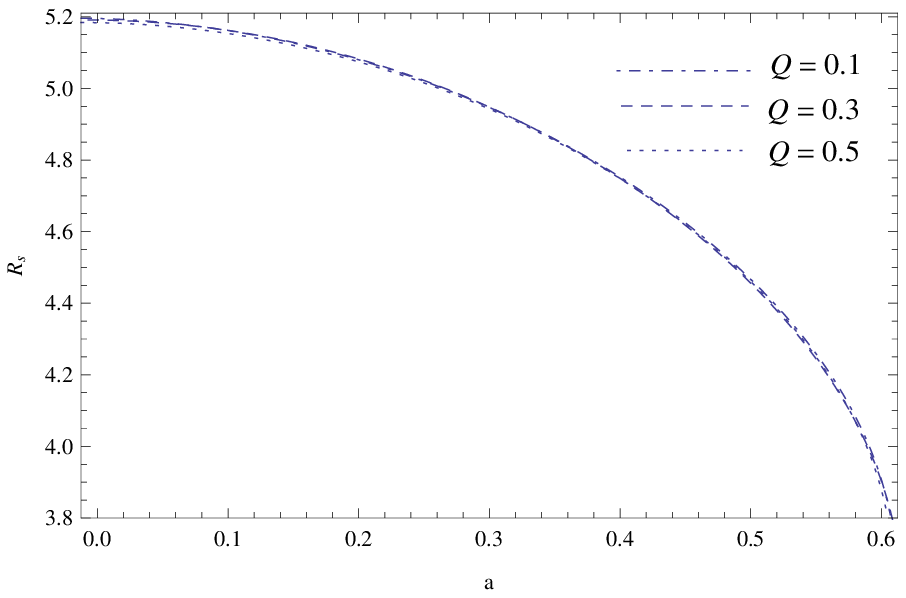,width=.40\linewidth}\epsfig{file=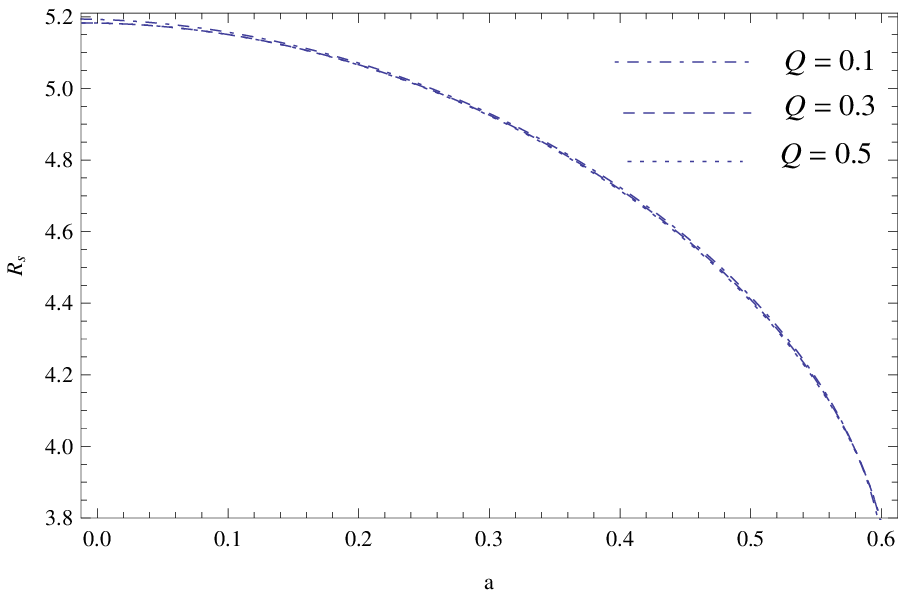,width=.40\linewidth}
\caption{The upper panel and lower panel shows radius of the shadow
and deformation parameter corresponding to $\pi/2$ (left), $\pi/3$
(right) and $\Theta=0.3$.}
\end{figure}

\section{Shadow of the BH Surrounded by Plasma}

In this section, we explore the effects of plasma on the shadow of
rotating charged BH in the background of NC geometry. For the
axially symmetric BH, the refraction index of the plasma is
$n=n(x^{i},w)$, where the photon frequency $w$ is measured by the
observer with velocity $u^{\nu}$. The effective energy of the photon
is $\hbar w=-p_{\nu}u^{\nu}$, ($p_{\nu}$ is the 4-momentum of the
photon). The relation between the refraction index and 4-momentum of
the photon is given as \cite{b}
\begin{equation}
n^2=1+\frac{p_{\nu}p^{\nu}}{p_{\sigma}u^{\sigma}},
\end{equation}
where $n=1$ in the absence of plasma. In order to find analytical
results, it is useful to write refraction index with specific plasma
frequency $w_{e}$ \cite{c}
\begin{equation}
n^2=1-\frac{w^2_{e}}{w^2}.
\end{equation}
The Hamilton-Jacobi equation for the photon around the BH surrounded
by a plasma has the following form \cite{c,b}
\begin{equation}
\frac{\partial S}{\partial
\lambda}=\frac{1}{2}[g^{\nu\sigma}p_{\nu}p_{\sigma}
-(n^2-1)(p_{0}\sqrt{-g^{00}})^2].
\end{equation}
This leads to the following equations of motion
\begin{eqnarray}\label{3m}
\Sigma\frac{\partial
t}{\partial\lambda}&=&a(L-an^2E\sin^2\theta)+\frac{r^2+a^2}{\Delta}[n^2E(r^2+a^2)-aL],
\\\label{6m}
\Sigma\frac{\partial
\phi}{\partial\lambda}&=&(L\csc^2\theta-aE)+\frac{a}{\Delta}[E(r^2+a^2)-aL],
\\\label{4m1}
\Sigma\frac{\partial r}{\partial\lambda}&=&\pm\sqrt{R_{p}},
\\\label{4m}
\Sigma\frac{\partial
\theta}{\partial\lambda}&=&\pm\sqrt{\Theta_{p}},
\end{eqnarray}
where
\begin{eqnarray}\label{141}
R_{p}&=&[E(r^2+a^2)-aL]^2+(n^2-1)(r^2+a^2)^2E^2
\\\nonumber
&-&\Delta[K+(L-aE)^2],\\\label{15}
\Theta_{p}&=&K+\cos^2\theta(a^2E^2-L^2\csc^2\theta)-a^2(n^2-1)E^2\sin^2\theta.
\end{eqnarray}

For the sake of simplicity, we chose a specific form of $w_{e}$ with
the assumption that the spacetime is filled with cold plasma
(non-magnetized) whose electron plasma frequency is a function of
the radial coordinate
$$w^2_{e}=\frac{4\pi e^2N(r)}{m_{e}},$$
where $e$ and $m_{e}$ represent the electron charge and mass
respectively, and $N(r)$ denotes the number density of the electrons
in plasma. Following \cite{c}, we consider $N(r)$ as the power law
density such that $N(r)=N_{0}/r^{h},~h\geq 0$ which yields
$w^2_{e}=k/r^{h}$, where $k$ is an arbitrary constant. Here we take
$h=1$. To study the effect of plasma on the shadow of charged
rotating NC BH, we follow the same procedure as in section
\textbf{2}. We find the parameters $\xi$ and $\eta$ of the form
\begin{eqnarray}
\xi&=&\frac{Y}{X}+\sqrt{\frac{Y^2}{X^2}-\frac{Z}{X}}, \\\nonumber
\eta&=&\frac{(r^2+a^2-a\xi)^2+(n^2-1)(r^2+a^2)^2}{\Delta}-(\xi-a)^2,
\end{eqnarray}
where
\begin{eqnarray}\nonumber
X&=&\frac{a^2}{\Delta},
\\\nonumber
Y&=&\frac{a}{\Delta(m(r)-r-m'(r)r-q(r)q'(r))}[m(r)a^2-m(r)r^2
\\\nonumber
&+&m'(r)r^3+m'(r)ra^2+q^2(r)r-q(r)q'(r) (r^2+a^2)],
\\\nonumber
Z&=&\frac{n^2(r^2+a^2)^2}{\Delta}+\frac{2rn^2(r)(r^2+a^2)-
n(r)n'(r)(r^2+a^2)^2}{(m(r)-r-m'(r)r-q(r)q'(r))}.
\end{eqnarray} The
celestial coordinates under the influence of plasma can be written
as
\begin{eqnarray}\nonumber
\alpha &=&-\xi\csc\theta_{0},\\\nonumber
\beta&=&\frac{\sqrt{\eta+a^2\cos^2\theta_{0}
-\xi^2\cot^2\theta_{0}-n^2a^2\sin^2\theta_{0}}}{n}.
\end{eqnarray}
Figure \textbf{5} indicates the effect of plasma for $a=0.4$,
$\theta=\frac{\pi}{2}$, $M=1$ and varying charge as well as $n$. We
observe that the shadow with plasma is greater than the case of
without plasma. The shadow increases with the increase of plasma
while increase in the value of charge tends to decrease the
deformation.
\begin{figure}\centering
\epsfig{file=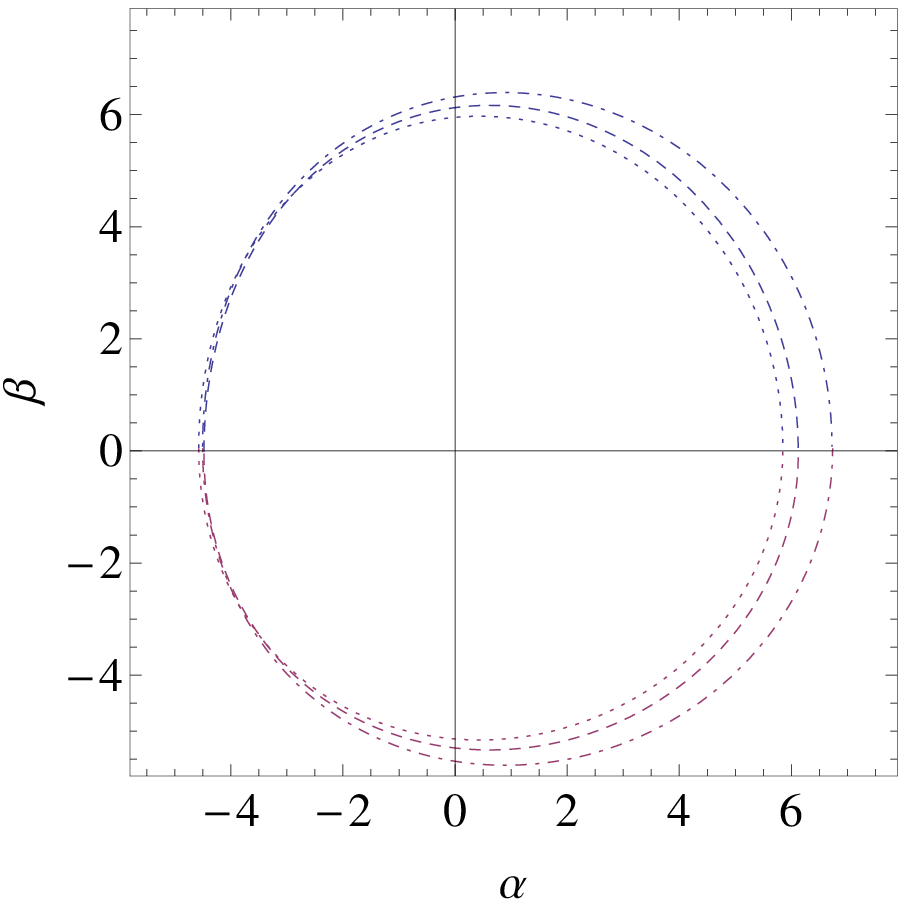,width=.35\linewidth}\epsfig{file=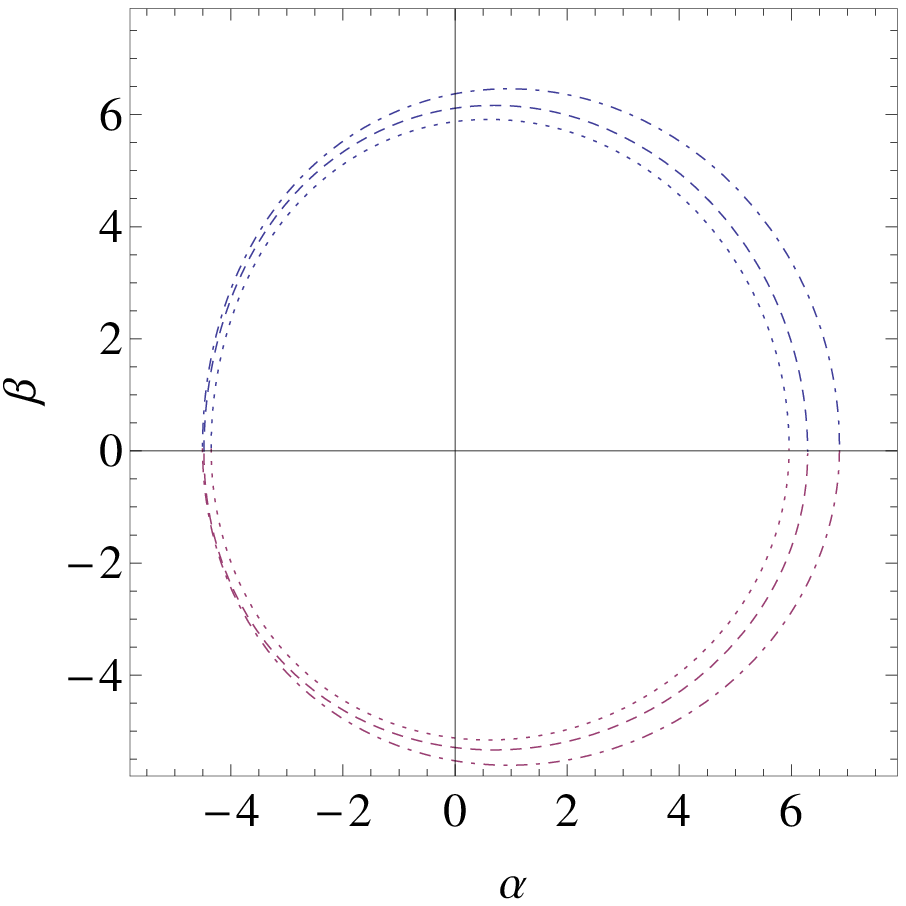,
width=.35\linewidth}\epsfig{file=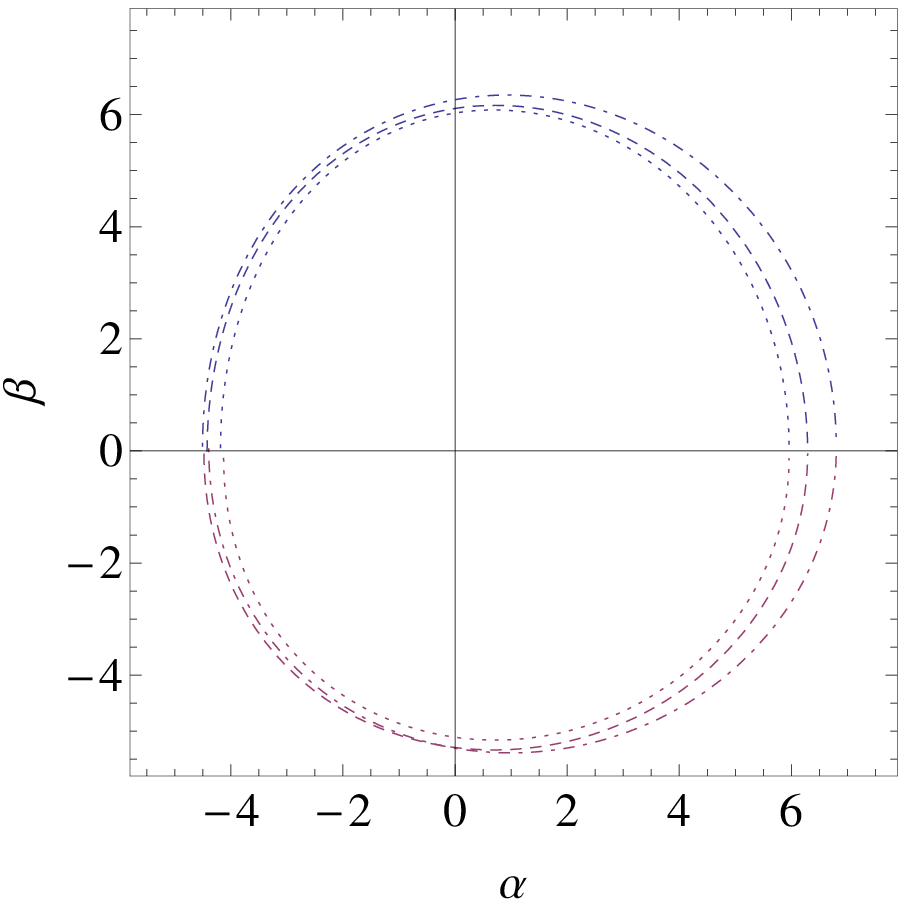,width=.35\linewidth}
\caption{Shadow of charged NC BH in the presence of plasma. Plots
from left to right correspond to $Q=0.1$, $Q=0.3$ and $Q=0.5$ where
dotted, dashed and dotted-dashed lines correspond to absence of
plasma, $k^2=0.1$ and $k^2=0.4$.}
\end{figure}

\section{Final Remarks}

The study of the apparent shapes of a BH shadow has been the subject
of interest for the last few years as it may be possible to observe
a BH in the center of galaxy in near future \cite{25}. The shadow
appears due to the effects of strong gravity near the BH, so it may
be helpful to explore its nature. For classical Kerr and Kerr-Newman
BHs, the shadow is affected by the spin, charge and the inclination
angle. Since the NC geometry has many applications such as resolving
the singularity problems in GR, it would be interesting to study the
effect of NC charge parameter on the shadow as well the particle
orbits. In this paper, we have examined the shadow cast by a charged
rotating NC BH. In order to study the effect of NC charge on the
silhouette of shadow, we have first obtained null geodesics and
analyzed the behavior of the effective potential. It is found that
the photons are stable for small values of radius but they approach
to maxima for the large radius which correspond to unstable orbits.
After reaching the maxima, $V_{eff}$ starts diverging.

We have used null geodesics to find the celestial coordinates
$\alpha$ and $\beta$ which describe the nature of the shadow. It is
found that the charged NC BH shadow is not only affected by NC
charge but also due to spin as well as the angle of inclination. We
observe that the silhouette of shadow changes its shape with respect
to varying charge parameter. The shape deviates from the circle for
the small charge while for the large NC charge, it approximately
maintains the shape of a circle. This behavior is almost similar to
the rotating uncharged NC BH \cite{20} (where the shape of the
circle is maintained with increase in the NC parameter) as well as
charged BHs \cite{v,v1}. The high values of spin as well as angle of
inclination deform the shape of the shadow. We have also discussed
the radius of shadow and deformation parameter for this BH. The
large value of NC charge corresponds to less deformation while for
rotating uncharged NC BH \cite{20} the distortion parameter
increases with NC parameter. This leads to an idea that NC charge
helps to maintain the shape of circle. The radius of the shadow
decreases as spin increases. Finally, we have examined the shadow of
charged rotating NC BH in the presence of plasma. The existence of
plasma increases the size of the shadow and deformation in the shape
reduces due to the presence of NC charge which is analogous to
\cite{v1}. We conclude that the presence of NC charge not only
affects the particle orbits but also affects shadow of the BH.

\vspace{0.25cm}

{\bf Acknowledgments}

\vspace{0.25cm}

This work has been supported by the \emph{Pakistan Academy of
Sciences Project}. We would like to thank Alejandro Cardenas for his
kind help to draw the figures.

\end{document}